\input{aipcheck}

\def\ltsima{$\; \buildrel < \over \sim \;$}
\def\simlt{\lower.5ex\hbox{\ltsima}} 

\def\gtsima{$\; \buildrel > \over \sim \;$}
\def\simgt{\lower.5ex\hbox{\gtsima}}

\documentclass[
    ,final            
  ]
  {aipproc}

\layoutstyle{6x9}

\begin{document}

\title{Simbol-X Hard X-ray Focusing Mirrors: Results Obtained During the Phase A Study}

\classification{95.55.-n; 95.55.Ka, 95.85.Nv}
\keywords      {X-ray, Instrumentation}

\author{G. Tagliaferri}{
  address={INAF-Osservatorio Astronomico di Brera, Via Bianchi 46, 23807 Merate, Italy}
}

\author{S. Basso}{
  address={INAF-Osservatorio Astronomico di Brera, Via Bianchi 46, 23807 Merate, Italy}
}

\author{G. Borghi}{
  address={Media Lario Technologies S.r.l., Localit\`a Pascolo, 23842 Bosisio Parini, Italy}
}

\author{W. Burkert}{
  address={Max-Planck-Institut f\"ur Extraterrestrische Physik, Giessenbachstr. 1, 85748 Garching, Germany}
}

\author{O. Citterio}{
  address={INAF-Osservatorio Astronomico di Brera, Via Bianchi 46, 23807 Merate, Italy},
 altaddress={Media Lario Technologies S.r.l., Localit\`a Pascolo, 23842 Bosisio Parini, Italy}
}

\author{M. Civitani}{
  address={INAF-Osservatorio Astronomico di Brera, Via Bianchi 46, 23807 Merate, Italy}
 }

\author{P. Conconi}{
  address={INAF-Osservatorio Astronomico di Brera, Via Bianchi 46, 23807 Merate, Italy}
}

\author{V. Cotroneo}{
  address={INAF-Osservatorio Astronomico di Brera, Via Bianchi 46, 23807 Merate, Italy}
}

\author{M. Freyberg}{
  address={Max-Planck-Institut f\"ur Extraterrestrische Physik, Giessenbachstr. 1, 85748 Garching, Germany}
}

\author{D. Garoli}{
  address={Media Lario Technologies S.r.l., Localit\`a Pascolo, 23842 Bosisio Parini, Italy}
}

\author{P. Gorenstein}{
  address={Harvard-Smithsonian Center for Astrophysics, 60 Garden Street, Cambridge, MA 02138, USA}
}

\author{G. Hartner}{
  address={Max-Planck-Institut f\"ur Extraterrestrische Physik, Giessenbachstr. 1, 85748 Garching, Germany}
}
  
\author{V. Mattarello}{
  address={Media Lario Technologies S.r.l., Localit\`a Pascolo, 23842 Bosisio Parini, Italy}
}

\author{A. Orlandi}{
  address={Media Lario Technologies S.r.l., Localit\`a Pascolo, 23842 Bosisio Parini, Italy}
}

\author{G. Pareschi}{
  address={INAF-Osservatorio Astronomico di Brera, Via Bianchi 46, 23807 Merate, Italy}
}

\author{S. Romaine}{
  address={Harvard-Smithsonian Center for Astrophysics, 60 Garden Street, Cambridge, MA 02138, USA}
}

\author{D. Spiga}{
  address={INAF-Osservatorio Astronomico di Brera, Via Bianchi 46, 23807 Merate, Italy}
}

\author{G. Valsecchi}{
  address={Media Lario Technologies S.r.l., Localit\`a Pascolo, 23842 Bosisio Parini, Italy}
}

\author{D. Vernani}{
  address={Media Lario Technologies S.r.l., Localit\`a Pascolo, 23842 Bosisio Parini, Italy}
}

\begin{abstract}
Simbol-X will push grazing incidence imaging up to 80 keV, providing a strong
improvement both in sensitivity and angular resolution compared to all instruments that have
operated so far above 10 keV. The superb hard X-ray imaging capability will be guaranteed
by a mirror module of 100 electroformed Nickel shells with a multilayer reflecting coating.
Here we will describe the technogical development and solutions adopted for the fabrication
of the mirror module, that must guarantee an Half Energy Width (HEW) better than 20 arcsec from 0.5 up to 30 keV and a
goal of 40 arcsec at 60 keV.  During the phase A, terminated at the end of 2008, we have developed
three engineering  models with two, two and three shells, respectively. The most critical aspects in
the development of the Simbol-X mirrors are {\it i)} the production of the 100 mandrels with very
good surface quality within the timeline of the mission, {\it ii)} the replication of shells that must be
very thin (a factor of 2 thinner than those of XMM-Newton) and still have very good image quality
up to 80 keV, {\it iii)} the development of an integration process that allows us to integrate these very
thin mirrors maintaining their intrinsic good image quality. The Phase A study has shown that we can
fabricate the mandrels with the needed quality and that we have developed a valid integration process.
The shells that we have produced so far have a quite good image quality, e.g. HEW$ \simlt 30$ arcsec
at 30 keV, and effective area. However, we still need to make some improvements to reach the
requirements. We will briefly present these results and discuss the possible improvements that we
will investigate during phase B.
\end{abstract}

\maketitle


\section{INTRODUCTION}

Thanks to the introduction of the focusing capability, the X-ray astronomy is now providing data
almost on all type of sources in the sky and the X-ray data are of common use in the astronomical
community almost as much as the optical, near-infrared and radio data. However, this is true only
in the soft-X-ray band, at energies between 0.1 and 10 keV. Above 10 keV, i.e. in the so called hard X-ray
band, so far we could use only passive collimators and coded mask. As a result only a few hundred
sources are known in the whole sky in the $10-100$ keV band. However, it is in this band that the X-ray background
has its peak, at $\sim 30$ keV and we still do not know which sources are making up most of this emission,
a question of cosmological importance. To discover them we need an instrument with much higher sensitivity
and imaging capability, to separate the various weak X-ray sources. This band is also very important for the study
of matter accretion around black holes and the particle acceleration mechanisms. For these reasons, there is a clear
need for a hard X-ray mission with a sensitivity and angular resolution similar to those of the current soft X-ray telescopes.
A hard X-ray focusing optics is needed to do this. With the emerging technology in mirror manufacturing, providing
that one can achieve a very long focal length and uses multilayer reflecting coatings, it is now possible to develop such a mission. 
In particular Simbol-X, a new mission currently under study by the French and Italian agencies will have the capability
to obtain X-ray images in the band $0.5-80$ keV with very good sensitivity and spatial resolution \cite{Ferrando}. 
The Simbol-X very long focal length of 20~m will be obtained by using a formation flying strategy; two satellites will carry
one the focusing mirrors and the other one the focal plane detector \cite{LaMarle}.  
While in order to meet the highly demanding image quality and effective area over the full $0.5 - 80$ keV energy band
a mirror module of 100 nested shells is foreseen \cite{Pareschi}. These shells will be Nickel replicated and
coated with a multilayer of more than 200 bilayers of Platinum and Carbon. Here we will describe the results that
we obtained during the Phase A study that has been completed in 2008, during which we fabricated three engineering models.

\section{MANDRELS, MIRRORS AND INTEGRATION PROCESS DEVELOPED DURING PHASE A}

\begin{figure}
  \includegraphics[height=.4\textheight]{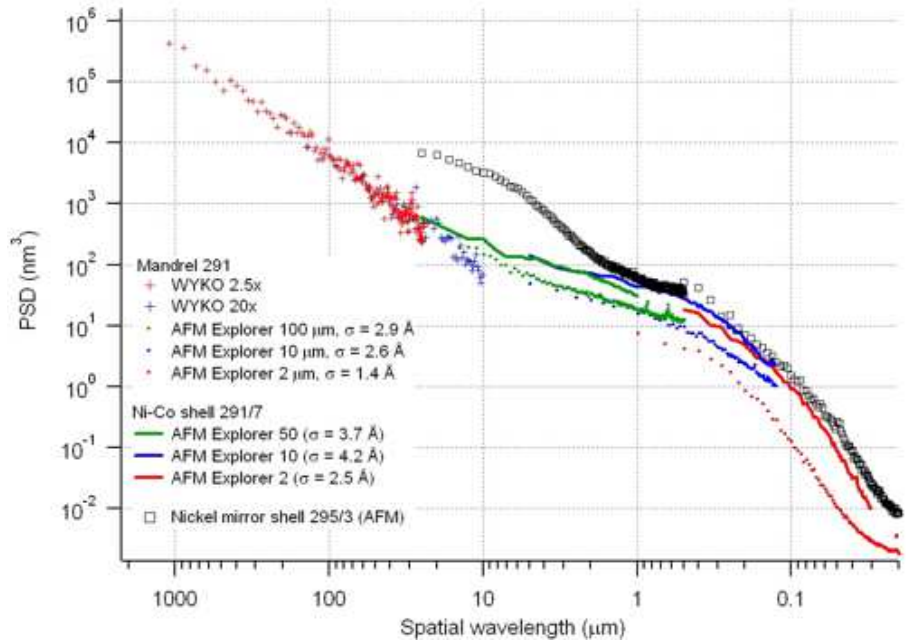}
  \caption{Power spectral density of the surface roughness as a function of the spatial wavelength for a mandrel
  developed in Phase A (crosses and points), a shell in Nickel (open squares) and a shell in Nickel-Cobalt both
  replicated during Phase A. Note that while the mandrel surface roughness is quite good, the replicated Ni-shell
  has a quite poor surface roughness. The shell in NiCo is much better up to a spatial wavelength of $\sim 2 \ \mu$m.
  Below this value also the surface roughness of the NiCo-shell needs to be improved. This problem will be further
  investigated during the Phase B.
  }
\end{figure}

The SIMBOL-X mirrors will be electroformed Ni shells with Wolter I profile. The adopted technology has been 
successfully used for the gold coated X-ray mirrors of the Beppo-SAX, XMM-Newton and Jet-X/Swift telescopes.
This technology has been developed and consolidated in the past two decades in Italy by the INAF-Brera 
Astronomical Observatory in collaboration with the Media Lario Technology company. For the Simbol-X mirrors, 
a few important modifications of the process will be implemented: 1) the use of multilayer reflecting coatings, 
allowing us to obtain a larger FOV and an operative range up to 80 keV and beyond; 2) the Ni walls will be a
factor of two thinner than the XMM mirror shells, to maintain the weight as low as possible. With respect to the 
first point, once the gold-coated Ni mirror shell has been replicated from the mandrel, the multilayer film will be
sputtered on the internal surface of the shell by using a two-targets linear DC magnetron sputtering system
\cite{Pareschi}. This process  has been  developed at the SAO-Center for Astrophysics (CfA) for monolithic
pseudo cylindrical shells \cite{Romaine} and now also at Media Lario where a multilayer coating facility
has been developed and installed. 

To replicate mirror shells with the requested image quality we must start from mandrels with a very good
surface roughness, even better than those produced for the XMM-Newton mirrors. In fact we need mandrels
with a surface roughness of the order of 1-2 \AA \ over the space wavelength up to a few microns.
During Phase A we have developed four mandrels, three using the classical approach already used for
the XMM-Newton ones (i.e. the mandrels are produced by grinding and then worked with a long lasting
lapping procedure to reach the requested surface roughness) and one using the Diamond Turning technique.
The latter approach produces mandrels with a better starting surface, therefore minimizing the lapping procedure.
The goal is to reduce the time for the lapping procedure by a factor of four.
The four mandrels have a diameter of 286, 291, 295 and 297 mm, respectively. For all of them the surface roughness
has been measured to be within the specification needed to meet the mission requirements, including the mandrel
made with the diamond turning (\#297). As an example in Fig. 1 we show the roughness profile that we
measured with different metrological systems  for mandrel \#291.

\begin{figure}
  \includegraphics[height=.19\textheight]{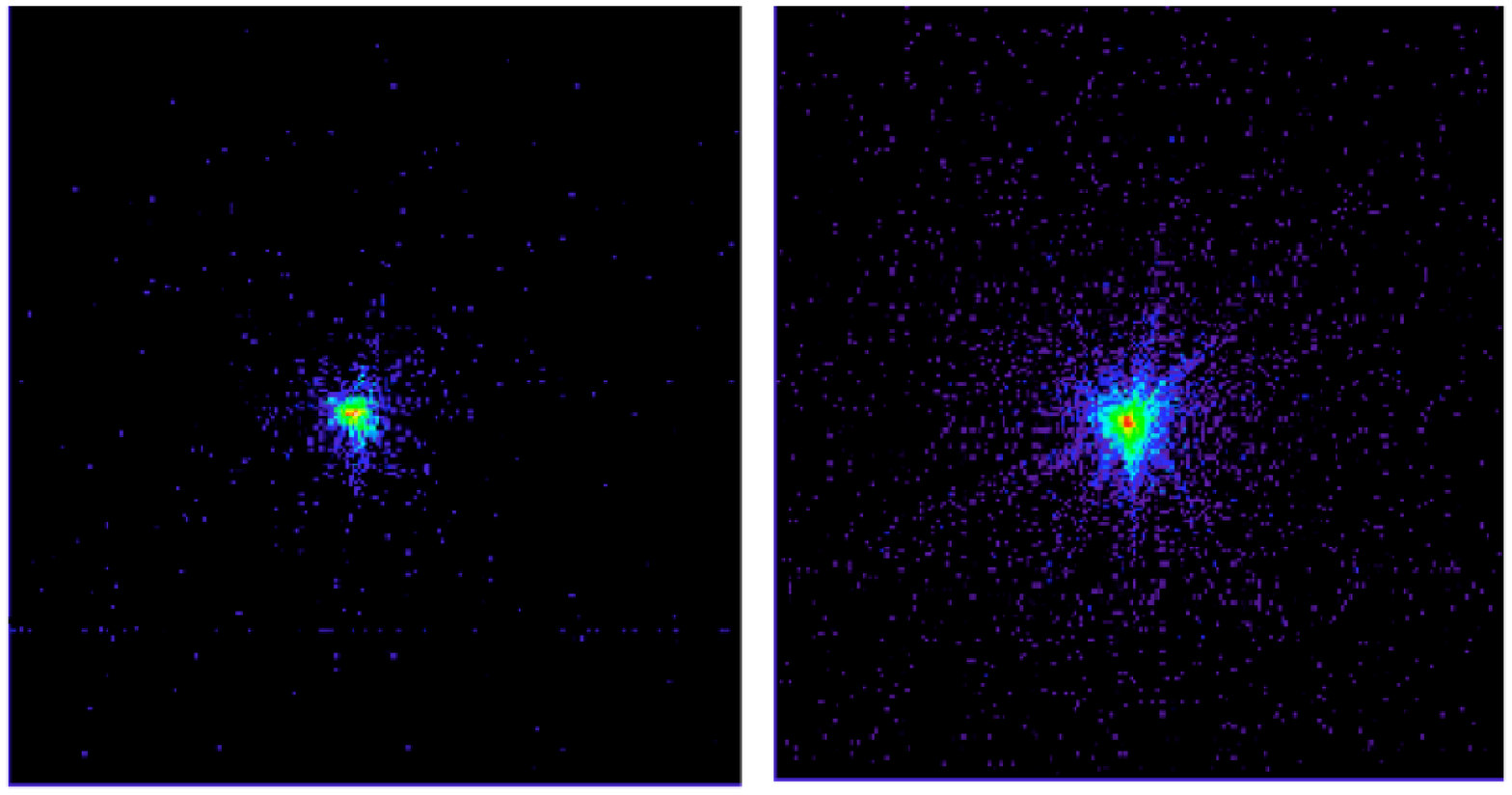}
  \includegraphics[height=.17\textheight]{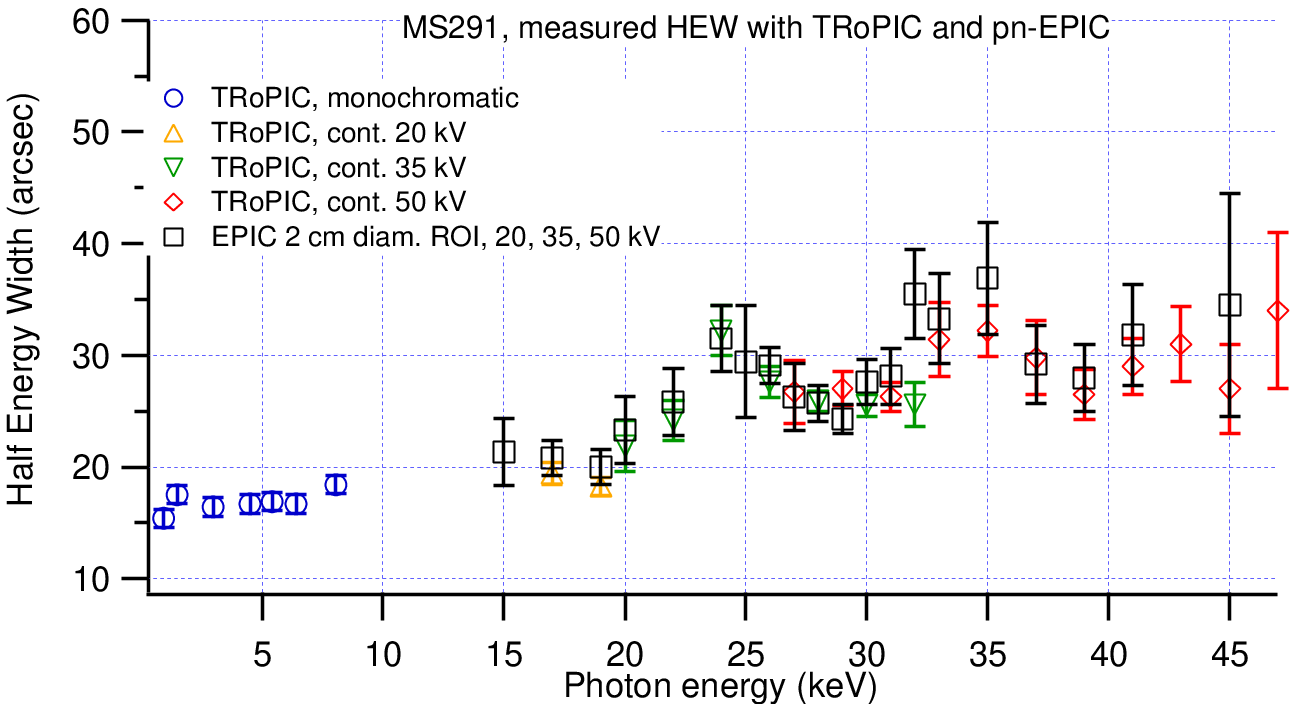}
  \caption{The image on the left has been taken at 0.93 keV at the Panter facility for the mirror shell 291
  integrated in the second engineering mirror module. The image in the middle has been taken for the
  same shell in the band 20-50 keV. As expected, at the higher energies the scattering is larger and the structure of the spider arms
can be noted, still  the image quality is extremely good and sharp up to 50 keV. This is also shown in the right panel, where we plot
the HEW of this shell as a function of energy. The HEW below 10 keV is about 18 arcsec, while that one at 30 keV
is `$\sim$25 arcsec,
not very far away from the requirements. See the text for more details.
}
\end{figure}

Using mandrels \#286, 291 and 295 we have replicated various mirror shells, using gold as release agent, and coated them
with a multilayer of 95 tungsten and silicon (W/Si) bilayers. We used W/Si because at the moment at the Panter facility, where
we performed the X-ray measurements, we can use X-ray only up to 50 keV and up to these energies W/Si and Pt/C have a similar
behaviour, but the W/Si is much cheaper. Also these mandrels and shells have a focal length of
10 m, once again because at the moment it is not possible to test mirrors with longer focal length at the Panter facility of MPE, in
Munich. It is already planned to upgrade
this facility in order to test mirrors with focal length of 20 m and up to 100 keV. We choose these diameters for the mandrels
in order to test the same incident angles, in turn related to the reflectivity, that will have the largest Simbol-X mirror shells with
a focal length of 20 m.

To test the integration procedure that we have developed for thin shells,
we integrated these shells in three different engineering models, the first two with two shells (\#295 \& 291) and the third one
with three shells (\#286, 291 \& 295) (we did not used mandrel \#297 because it was completed only at the end of this phase).
This procedure is based on the use of stiffening rings that are put around the shell and are removed after the shell has been glued
on the spider. We showed that this procedure meets the Simbol-X requirements (see Basso et al. these proceedings).  For the first engineering
model the multilayer deposition has been performed at the CfA, while for the other two this has been performed at Media Lario.

In Fig. 2, left and middle panels, we show two images taken at the Panter facility at 0.93 keV and in the band 30-50 keV with the shell \#291 integrated in
the second engineering model. As expected, at the higher energies the scattering is larger and the structure of the spider arms
can be noted, still  the image quality is extremely good and sharp up to 50 keV. This is also shown in Fig. 2 right panel, where we plot
the HEW of this shell as a function of energy. The HEW below 10 keV is about 18 arcsec, while that one at 30 keV
is $\sim$ 25 arcsec,
not very far away from the requirements.  The very slow increase of the HEW from 1 to 45 keV, due to X-ray scattering superposing
to the mirror figure error (responsible for the low-energy HEW), is probably due to a higher mirror surface roughness, with respect
to that one required by the mission. In fact in Fig. 1 we can see that while the mandrel surface  is very good, that one of
the replicated Ni shell is quite worse, in particular at the middle spatial wavelengths which are the ones influencing more the scattering.
This needs to be improved. But we need to improve also at the higher spatial wavelengths and have a better mirror figure error and,
therefore, a better HEW also below 10 keV. To this end, Media Lario started a new program in order to electro-form shells in Nickel-Cobalt
(NiCo), a metal alloy that has better structural performances than the pure Nickel, in particular it has a higher yield. Therefore these shells
should be stiffer and more resistant to plastic deformation during the release process.
The third engineering model has been assembled with three shells already made in NiCo. These shells, when measured on our optical
bench, have shown an improvement of their intrinsic optical quality, therefore we would have expected an improvement of the HEW
as measured in the X-ray at the Panter facility, at least at lower energies. However, due to an error during the integration process,
some bolts were tightened too much, we degraded
somewhat the image quality of the integrated shells. Therefore the values that we measured at the Panter facility were only slightly better.
In summary, we are confident that the use of the NiCo alloy will improve the intrinsic image quality of the
mirror shells, but we do not have X-ray measurements to prove this, yet. 

However, from Fig. 1 we can already see that the NiCo shells will expectedly have better optical performances. Note how the surface roughness
measured for the NiCo shell (color-continued lines) is much better than that one of the pure Ni-shell (open square black points). This is
particularly true in the $30 < l < 2 \ \mu$m spatial wavelength range. However, below $1 \ \mu$m the surface roughness of the two shells
are very similar and quite worse with respect to the surface roughness of the mandrel. Therefore, we still need to improve over the
release process in order to have better mirror roughness also at very low spatial wavelength, that have  a stronger impact on the mirror
performances at higher energies.
This is reflected also by the measurements of the effective area as a function of energy. In Fig. 3 we compare the
effective area measured
at the Panter facility for the Ni-shell \#291 and the NiCo-shell \#286, with the theoretical expectations. While for the Ni-shell the
measured values are $\sim$ 10\% lower than the predicted one below 10 kev, $\sim$ 30\% lower between 15-20 keV and
$\sim$ 50\% lower at 30 keV;
for the NiCo-shell the measured values below 10 keV are in perfect agreement with the theoretical ones. From 15 to 20 keV the NiCo-shell
is still better than the Ni-shell, although somewhat below the theoretical values. While above 25 keV also the NiCo-shell is well below the
theoretical values. In fact, for the niCo-shell the measured values are more consistent with the theoretical model assuming a 8 \AA \ roughness,
instead of the requested 4 \AA.

\begin{figure}
  \includegraphics[height=.17\textheight]{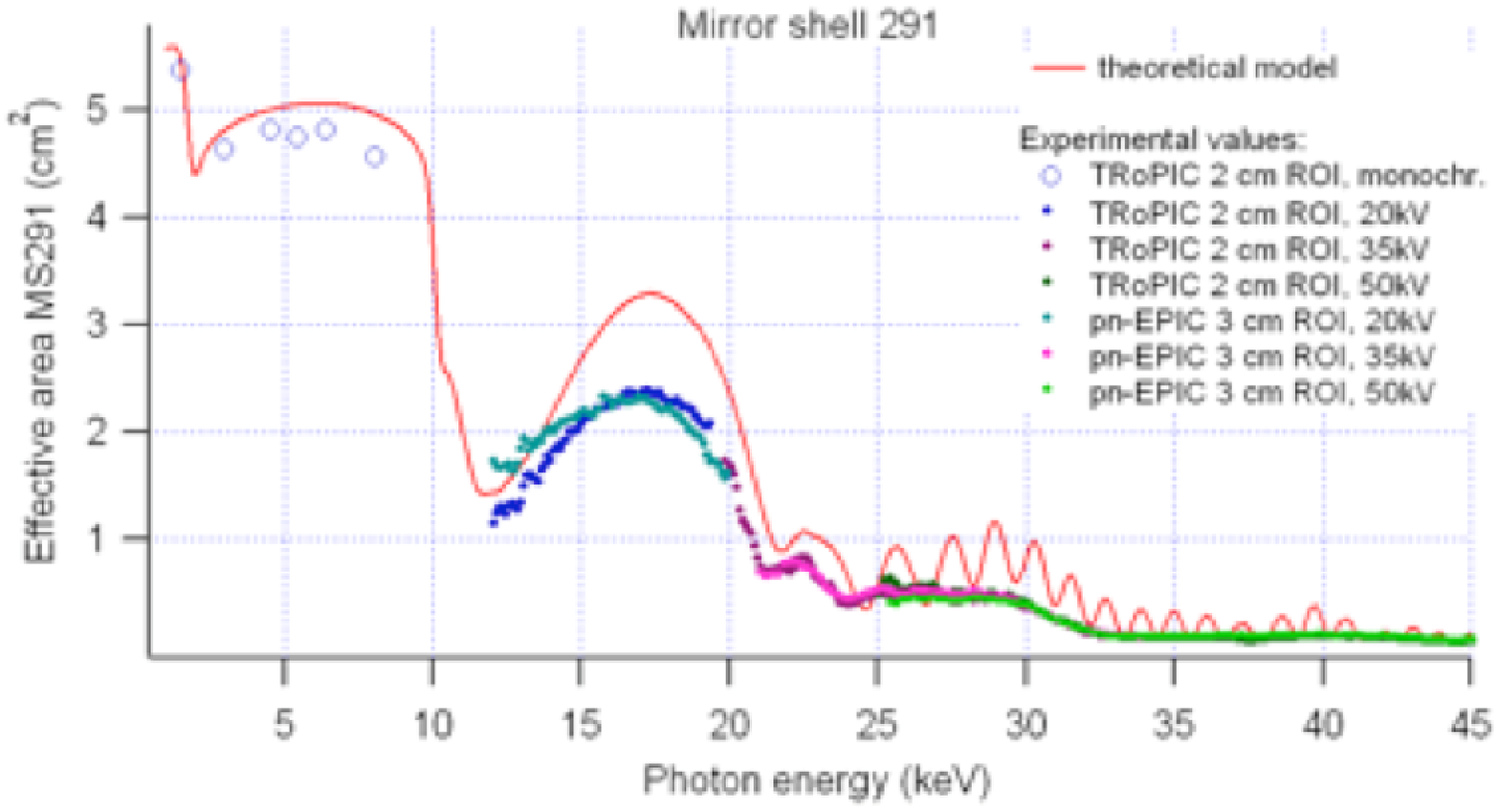}
  \includegraphics[height=.17\textheight]{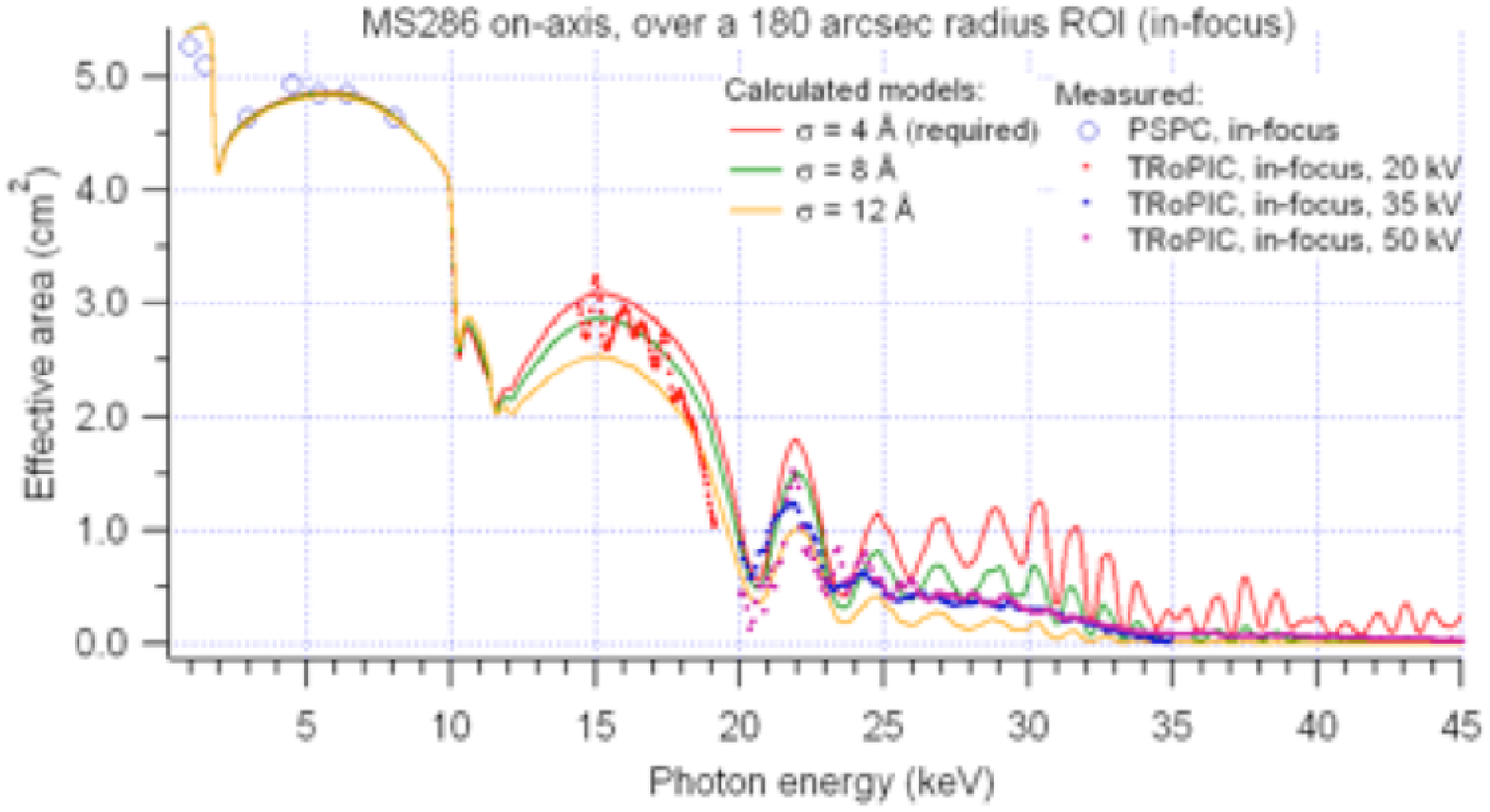}
  \caption{In the left panel we plot the on-axis effective areas for the mirror shell 291 in Ni from
the second engineering model as function of energy. The measured values are compared with the theoretical ones
assuming a mirror surface roughness of 4 \AA, requested by the mission. Clearly the measured values are below the expected ones.
In the right panel we have the same plot for the mirror shell 286 in NiCo from the second engineering model. The data
are compared with the theoretical values assuming three different surface roughness and are better represented by a model
with a surface roughness of 8 \AA. Clearly the NiCo shell is better, but we still need to improve the response at higher energies,
this problem will be further investigated in Phase B. For further details see the text. Note that the wiggles that are present above
20-25 keV in the theoretical values will not be present in the effective area of the final mirror module, that will have a smoother behaviour.
  }
\end{figure}

\section{CONCLUSIONS}

The phase A study for the realisation of the Simbol-x mirror module has been completed,
showing that the requirements of the mission can be meet. However, some further
technological developments are necessary and will be investigated during phase B.

The design of the mandrels has been optimized to reach very good longitudinal profiles
and surface roughness able to provide an intrinsic HEW of about 7 arcsec.
We showed that diamond turning technique can provide  mandrels that
have the surface quality that is needed by the mission 

The HEW of each shells was not degraded in a significant way by the
integration: the concept of the integration of thin shells with the 
stiffening rings works very well. However, we need a better control on
the behavior of the temporary structure during the integration process,
in order to avoid mistakes as the one that occurred during the integration of
the third engineering model.

We need a better control on the shell roughness degradation introduced by
the electroforming and release processes, in particular at the spatial wavelength
below $\simlt 1 \ \mu$m. This aspect will be investigated and improved during Phase B.


\begin{theacknowledgments}
This work has been financed by the Italian Space Agency as part of the industrial contract 
to carry out the Italian Simbol-X Phase A study led by Thales-AleniaSpace-Torino.
\end{theacknowledgments}

\bibliographystyle{aipproc}   

\end{document}